\documentclass[reprint,amsmath,amssymb,aps,pra]{revtex4-2}

\usepackage{float}
\usepackage{xcolor}
\usepackage{soul}
\usepackage{braket}
\usepackage{tikz}
\usetikzlibrary{quantikz2,decorations.pathmorphing}
\usetikzlibrary{arrows.meta}
\usepackage[caption=false]{subfig}
\usepackage{ragged2e}
\DeclareCaptionJustification{justified}{\justifying}
\usepackage[outdir=./]{epstopdf}

\usepackage{multirow}
\usepackage{array}

\usepackage{graphicx}
\usepackage{dcolumn}
\usepackage{bm}

\newcommand{\PB}[1]{\textcolor{black}{#1}}
\newcommand{\PBs}[1]{}
\newcommand{\PBc}[1]{}

\begin{document}

\title{Intractability of Witnessing \PB{an }Entangled Measurement\PBs{s} Device Independently}

\author{Peter Bierhorst} 
\email{plbierho@uno.edu}

\affiliation{Mathematics Department, University of New Orleans, Louisiana, USA}

\date{\today}

\begin{abstract}
Protocols have been previously proposed to certify that a measurement is entangled in a fully device-independent manner. Here, I provide models for these protocols in which the claimed measurement is not entangled, and demonstrate it is always possible to displace entanglement from a given measurement to the measured state for a general class of device-independent scenarios. This indicates that no black-box measurement scenario requires a specific chosen measurement to be entangled in order to replicate its behavior, which is relevant to our fundamental understanding of this phenomenon and how to witness it.
\end{abstract}

\maketitle

Entangled quantum states enable a range of fascinating nonclassical phenomena \cite{BBP}.  Quantum mechanics also describes entangled \textit{measurements}, which are arguably less well understood \cite{gisin:2019} despite their central role in fundamental tasks such as teleportation and entanglement distribution. The mathematical definition of an entangled measurement -- a quantum measurement acting on a state space of the form  of $\mathcal H_A \otimes \mathcal H_B$ where at least one measurement operator does not admit a separable form -- is precise, but not illuminating. A fuller understanding may be sought by studying experimental scenarios in which entangled measurements enable certain behaviors that might not otherwise be possible. In this vein, a \textit{device-independent} certificate of an entangled measurement infers its presence directly from experimental statistics without relying on detailed assumptions modeling experimental devices. 
In addition to enhancing our fundamental understanding of entangled measurements, device-independent protocols have applications such as certifying that quantum hardware is functioning properly even in an untrusted provider scenario. 

Device-independent protocols spatially separate measuring parties to ensure that the underlying measured states factor in a certain way; then, if certain experimental statistics are observed, strong conclusions can be drawn about the specific states and measurements that must be present. Protocols certifying the presence of an entangled measurement often invoke additional device-dependent assumptions, such as an assumption that two sources of entangled quantum states operate independently \cite{renou:2018}, or that all quantum sources are at most bi-partite \cite{bierhorst:2023}, or that the dimensionality of the measured systems is known \cite{vertesi:2011}. 
Some protocols however have claimed to certify that a measurement is entangled in a completely device-independent manner \cite{rabello:2011,cong:2017,bancal:2015,bancal:2018}; that is, with no assumptions about the states being measured beyond that which can be guaranteed through separation of measuring devices alone.

Here I re-examine fully device-independent scenarios claiming to require an entangled measurement, and provide alternative models in which the measurement in question is not entangled (i.e., it is separable). The root issue is that device-independent guarantees on states and measurements are true only \textit{up to local isometry}. This enables a key trick of displacing entanglement from the measurement to the measured state, though this can require some subtle and/or complicated manipulations depending on the scenario, and may -- depending on interpretation of subsystems -- require replacing other measurements in the scenario that were not previously entangled with entangled ones. Nonetheless, the highlighted measurement can always be taken to be separable, and in particular realized as independent computational basis measurements of physically distinct qubits, with (at most) a potential classical post-processing step of the outputs of these computational basis measurements. The models are compatible with the strictest enforcement of device-independence through space-like separation of measuring events.

Moving beyond tailored models for specific scenarios, I prove a general impossibility result by exploiting a connection to \textit{localization protocols} for performing entangled measurements in a distributed manner. Suitably modifying and adapting a localization protocol due to Vaidman \cite{vaidman:2003} enables construction of a procedure for removing an entangled measurement from any device-independent scenario. The procedure applies to a general class of scenarios allowing for quantum information to be transported to different locations and measured at different times. Under a stringent definition of entangled measurements whereby no measurement on a single higher-dimensional qudit is considered entangled over subsystems, the procedure can be iterated to remove all entangled measurements from a measurement scenario; this stringent definition is however not required for removing just one entangled measurement.

\medskip

\begin{figure}
\hspace{-2mm}\includegraphics[scale=0.65]{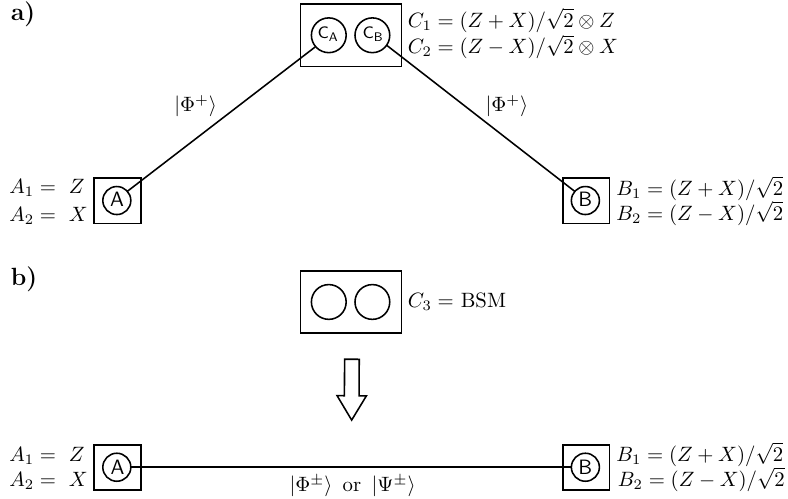}
\caption{The scheme of Rabelo \textit{et al}, in which Charlie ($\mathsf C = \mathsf {C_A}$ and $\mathsf {C_B}$) either a) performs two parallel CHSH-Bell tests with $\mathsf A$ and $\mathsf B$, or b) a Bell state measurement (BSM) which swaps an entangled state to $\mathsf A$ and $\mathsf B$. }\label{f:rabeloscheme}
\end{figure}

\noindent \textit{Rabelo et al.~protocol.} The protocol of Rabelo \textit{et al.}~\cite{rabello:2011} was the first to claim a fully device-independent certification of an entangled measurement. Ref.~\cite{rabello:2011} describes a tripartite scenario comprising three parties Alice, Bob, and Charlie -- respectively $\mathsf A$, $\mathsf B$, and $\mathsf C=(\mathsf{C_A},\mathsf{C_B})$ in Figure \ref{f:rabeloscheme}.  Alice and Bob each toggle between two binary-outcome measurements while Charlie has three four-outcome measurement settings $C_1$, $C_2$, and $C_3$. Charlie uses measurements $C_1$ and $C_2$ to perform two parallel CHSH-Bell tests with $\mathsf A$ and $\mathsf B$ on shared maximally entangled Bell states, depicted in Fig.~1(a). Alternatively, on measurement setting $C_3$, Charlie performs an entangled Bell state measurement (BSM) which swaps to Alice and Bob one of the four canonical Bell states
\begin{equation*}
\ket {\Phi^{\pm}} =  (\ket{00}\pm\ket{11})/2 \qquad \ket {\Psi^{\pm}} =  (\ket{01}\pm\ket{10})/2
\end{equation*}
depending on Charlie's outcome; see Fig.~1(b). Now Alice and Bob's CHSH-Bell measurements result in a maximal violation of the CHSH inequality -- or one of its symmetries, depending on which Bell state results from the swap -- with each other. Ref.~\cite{rabello:2011} argues, citing the strict restrictions that maximal violations of CHSH inequalities impose upon measured states via self-testing \cite{supic:2020}, that observing the behavior of the scheme in Figure \ref{f:rabeloscheme} device-independently certifies that the measurement on setting $C_3$ is entangled.

\begin{figure}
\includegraphics{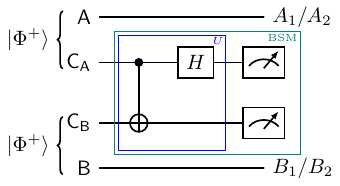}
\caption{Circuit diagram for measurement setting $C_3$ in the Rabelo \textit{et al.}~scheme, corresponding to Fig.~\ref{f:rabeloscheme} (b). As the Bell state measurement at $\mathsf C = (\mathsf{C_A},\mathsf{C_B})$ (BSM, outer box) can be implemented with a unitary $U$ (inner box, see also equation \eqref{e:Udef}) followed by two independent computational basis measurements (meters), we can replace it with an un-entangled measurement by absorbing $U$ into the measured state.}\label{f:BSM}
\end{figure}

The experiment can however be modeled with an unentangled measurement for $C_3$. Recalling that a BSM can be implemented as in Figure \ref{f:BSM} with a CNOT gate and a Hadamard ($H$) gate followed by independent measurements in the computational $\ket 0 \!/\! \ket 1$ basis (\cite{NC}, Sec.~1.3.7), one can displace the application of the unitaries to the measured state: that is, apply to $\ket {\Phi^+}\otimes \ket {\Phi^+}$ the operator $I_A \otimes U_{\mathsf{C_AC_B}}  \otimes I_B$, where
\begin{equation}\label{e:Udef}
U_{\mathsf{C_AC_B}} = (H_{\mathsf{C_A}} \otimes I_{\mathsf{C_B} })(CNOT_{\mathsf{C_AC_B}} ),
\end{equation}
to get a 4-qubit entangled state $\ket \Lambda$ given by
\begin{equation*}
\ket \Lambda = \sum_{i=0}^1 \left[\ket{0i00}+\ket{0i11}+(-1)^i(\ket{1i10}+\ket{1i01})\right]/2\sqrt{2}.
\end{equation*}
This state can be prepared at far remove from the measurement events and independently of the selection of measurement settings. Then in the alternative model, the measured tripartite state is $\ket \Lambda$, the four-outcome measurement $C_3$ is a product measurement measuring the middle two qubits independently in the computational basis, and the measurement operators for $C_1$ and $C_2$ consist of the application of $U^\dagger $, which restores $\ket \Lambda$ to $\ket {\Phi^+}\otimes \ket {\Phi^+}$, followed by the CHSH measurements of the original model in Figure \ref{f:rabeloscheme} (a). 
This is possible, and compatible with the reasoning in Ref.~\cite{rabello:2011}, because the CHSH measurements $C_1$ and $C_2$ self-test Bell states only \textit{up to local isometry}, and $U^\dagger$ is precisely the local isometry that can realize the pair of Bell states while dis-entangling $\ket \Lambda$ across the $C_A$/$C_B$ partition. This loophole also allows one to replicate a similar behavior described in Ref.~\cite{cong:2017} in the vicinity of Eq.~(4) therein, such that the measurement claimed to be certified entangled -- setting $y=4$ in that work -- is not (though note a more central aim of Ref.~\cite{cong:2017} was to certify a lower bound on Hilbert space dimension of measured states, which is not invalidated by an alternate model removing an entangled measurement).

For the above model where $C_3$ comprises two independent computational basis measurements, the reasoning of Ref.~\cite{rabello:2011} leads to an interpretation of $C_3$ an entangled measurement on \textit{virtual} qubits encoded in the Bell states across the real physical qubits comprising the second and third qubits of the state $\ket \Lambda$. This interpretation is inbued with more meaning by a context where these virtual qubits represent degrees of freedom that are effectively entangled as components of Bell states shared with systems $\mathsf A$ and $\mathsf B$, as witnessed by the maximal CHSH violations on measurements $C_1$ and $C_2$. Nonetheless, the measurement of the actual physical qubits on setting $C_3$ remains two independent computational basis measurements; if we imagine qubits two and three as (say) polarization states of separate photons arriving into Charlie's measuring box through separate wires, the photons never interact when measurement setting $C_3$ is selected: they are routed to separate locations within the measurement box where they are each measured separately. It is reasonable to interpret this as not comprising an entangled measurement. 

Notably, the above model still contains entangled measurements -- they have just been displaced from $C_3$ to $C_1$ and $C_2$, as $U^\dagger$ is entangling. Moreover, a key vulnerability enabling the above model is the co-location of the $C_A$ and $C_B$ measurements on settings $C_1$ and $C_2$. Indeed, this allows a more trivial way to dispute that the Rabelo \textit{et al.}~scenario requires an entangled measurement: one can just relabel the qubit basis states of the form $\ket {ij}_{\mathsf C_A \mathsf{C_B}}$ for $i, j \in \{0,1\}$ as a four-level system or ququart $\ket k_{\mathsf C}$ with $k \in \{0,1,2,3\}$:
\begin{equation}\label{e:ququart}
\ket{00} \to \ket 0 \quad
\ket{01} \to \ket 1 \quad
\ket{10} \to \ket 2 \quad
\ket{11} \to \ket 3. 
\end{equation}
All three measurements $C_1$,  $C_2$ and $C_3$ can then be modeled as various unentangled measurements on this four-dimensional system. The assertion that here \textit{no} measurement is unentangled, however, is not as compelling -- if the four-level system admits any sort of bipartition it appears at least one of the measurements would need be entangled over it.

\medskip

\noindent \textit{Bancal et al.~protocol.} One can attempt to address the vulnerability above with a protocol that operates in stages: in some rounds, the quantum particles are measured separately at different locations, while in other rounds, the quantum particles are transported to a mutual location where they can be subjected to an entangled measurement. Such a spacetime-ordered experiment can  be considered fully device independent, enforcing a state space of the form $\mathcal H_\mathsf{A} \otimes \mathcal H_\mathsf{C_A} \otimes \mathcal H_\mathsf{C_B} \otimes \mathcal H_\mathsf{B}$ in which  $\mathcal H_\mathsf{C_A}$ and $ \mathcal H_\mathsf{C_B}$ are measured separately in some experimental trials and jointly in others.  More recent protocols claiming fully device-independent certification of an entangled measurement -- see \cite{bancal:2015} Section VB, with further development in \cite{bancal:2018} (see Fig.~2 therein) -- do not explicitly describe the above loophole in the Rabelo \textit{et al.}~\cite{rabello:2011} scheme, but are careful to describe refined scenarios that are quadripartite in some rounds in which $C_A$ and $C_B$ are measured separately, and tripartite in others where $C_A$ and $C_B$ are brought together for joint measurement.

\begin{figure}
\hspace{0cm}\includegraphics[scale=.85]{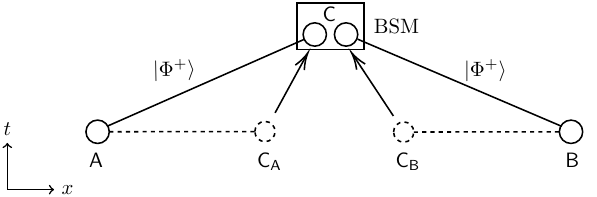}
\caption{Spacetime layout for the scheme of Bancal \textit{et al.} \cite{bancal:2018}. In various rounds, either $\mathsf {C_A}$ and $\mathsf {C_B}$ perform measurements self-testing Bell states shared with $\mathsf A$ and $\mathsf B$ (dashed lines), or they physically send their qubits to $\mathsf C$ for an entangled BSM.}\label{f:bancalscheme}
\end{figure}

Let us explore how the approaches of Refs.~\cite{bancal:2015,bancal:2018} can be formalized in the strongest device-independent implementation, whereby distinct measurement events are spacelike separated and we assume it is possible to select measurement settings (including instructions to transport quantum information to a different location) with free random choices that are uncorrelated with measured states. This will impose the strictest possible conditions upon alternative underlying models for the experiment. It also allows a natural way to classify which components of an experiment comprise a given measurement: the manipulations that occur after the measurement setting is provided, leading up to recording of the measurement outcome as classical information.

Let us arrange experimental elements as in Figure \ref{f:bancalscheme}, positioning the spacetime event of the entangled measurement at $\mathsf C$ in the forward lightcone of both $\mathsf{C_A}$ and $\mathsf{C_B}$ while all other spacetime event pairs are spacelike separated. The measurement scenario is otherwise similar to that of Ref.~\cite{rabello:2011}, toggling between rounds that self-test the pair of Bell states at $\mathsf{C_A}$/$\mathsf{C_B}$ and rounds that implement the BSM at $\mathsf C$. (Ref.~\cite{bancal:2018} employs slightly different pairs of Bloch sphere measurements than \cite{rabello:2011} at $\mathsf{C_A}$/$\mathsf{C_B}$ for self-testing, but this does not matter for the model discussed below.) The arrangement in Figure \ref{f:bancalscheme} precludes use of the earlier model, because now-separated $\mathsf {C_A}$ and $\mathsf{C_B}$ cannot apply $U^\dagger$ to disentangle $\ket \Lambda$.

\begin{figure}
\hspace{5.29cm}\scalebox{0.8}{
\begin{tabular}{cccccc}
&&\multicolumn{4}{c}{$\mathsf{C_B}$ BSM}\\
&& $\Phi^+$& $\Phi^-$& $\Psi^+$& $\Psi^-$\\ \cline{3-6}
&\multicolumn{1}{c|}{$\Phi^+$} &  $\Phi^+$& $\Phi^-$& $\Psi^+$& $\Psi^-$\\ 
$\mathsf{C_A}$&\multicolumn{1}{c|}{$\Phi^-$} &  $\Phi^-$& $\Phi^+$& $\Psi^-$& $\Psi^+$\\ 
BSM&\multicolumn{1}{c|}{$\Psi^+$} &  $\Psi^+$&$\Psi^-$& $\Phi^+$&$\Phi^-$\\ 
&\multicolumn{1}{c|}{$\Psi^-$}& $\Psi^-$ & $\Psi^+$& $\Phi^-$&$\Phi^+$\\
&&\multicolumn{4}{r}{$\mathsf{AB}$ state}\\
\end{tabular}
}

\vspace{-1cm}

\hspace{-3mm}\includegraphics[scale=.7]{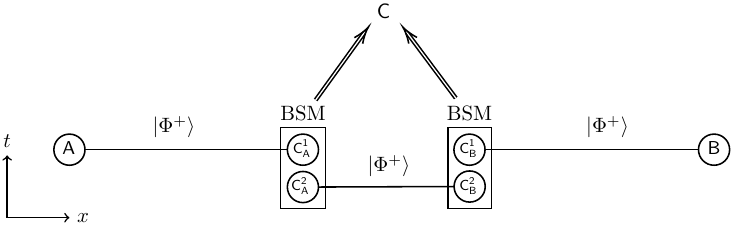} 
\caption{A scheme replicating the behavior of Figure \ref{f:bancalscheme} in which only classical information is transported and no quantum measurement is performed at event $\mathsf C$. The table indicates the state of $\mathsf A$ and $\mathsf B$ conditioned on the BSM outcomes.}\label{f:pironiostyle}
\end{figure}

To replicate the behavior in Figure \ref{f:bancalscheme} without an entangled measurement at $\mathsf C$, first consider a preliminary scheme depicted in Figure \ref{f:pironiostyle}, which exploits an ancillary Bell pair $\ket {\Phi^+}_{\mathsf{C_A^2C_B^2}}$ shared between $\mathsf{C_A}$ and $\mathsf{C_B}$. The goal of this scheme is to replace quantum channels with classical ones, reminiscent of the attack of Lobo, Pauwels, and Pironio (\cite{lobo:2024}, Sec.~2.1) on a protocol of Ref.~\cite{chaturvedi:2024}. Here in Fig.~\ref{f:pironiostyle}, when $\mathsf{C_A}$ and $\mathsf{C_B}$ receive instructions to implement their (unentangled, local) self-testing CHSH-type measurements, they ignore the ancillary Bell pair and directly measure the qubits entangled with $\mathsf A$ and $\mathsf B$. In trials when they receive alternate instructions to send their qubits to $\mathsf C$, they instead each locally perform a joint BSM on their pair of qubits -- now using $\ket {\Phi^+}_{\mathsf{C_A^2C_B^2}}$ -- and transmit the classical outcomes to $\mathsf C$. The two BSMs project the state of $\mathsf A$ and $\mathsf B$ into one of the four Bell states according to the table in Figure \ref{f:pironiostyle}, so $\mathsf C$'s classical information allows him to report an outcome correctly reflecting this state. While emphasizing that the overall effect of $\mathsf C$'s measurement on the  $\mathsf{C^1_A}\otimes \mathsf{C^1_B}$ state across all three locations is still entangled -- this is not yet the unentangled model -- the measurement is performed in a distributed manner such that the physical location of $\mathsf C$ need not contain any quantum-capable hardware, which is problematic in an untrusted provider scenario attempting to certify the performance of a BSM at $\mathsf C$. 

One removes the entangled measurement completely by modifying the model of Figure \ref{f:pironiostyle} to displace the entanglement of the two BSMs to the measured state. Specifically, replace the state ${\ket {\Phi^+}_{\mathsf {AC^1_A}}\otimes \ket {\Phi^+}_{\mathsf {C^2_AC^2_B}}\otimes \ket {\Phi^+}_{\mathsf {C^1_BB}}}$ with the six-qubit state $\ket \Xi$ defined as  
\begin{equation*}
(I_{\mathsf A} \otimes U_{\mathsf{C^1_AC^2_A}}\otimes U_{\mathsf{C^2_BC^1_B}}\otimes I_{\mathsf B})\ket {\Phi^+}_{\mathsf {AC^1_A}}\otimes \ket {\Phi^+}_{\mathsf {C^2_AC^2_B}}\otimes \ket {\Phi^+}_{\mathsf {C^1_BB}}
\end{equation*}
where $U$ is the two-qubit unitary defined in \eqref{e:Udef}. Now when $\mathsf{C_A}$ and $\mathsf{C_B}$ are tasked to perform  CHSH/self-test-type measurements, they each first perform $U^\dagger$ locally to recover the original state of $\mathsf{AC^1_A}$ or $\mathsf{C^1_BB}$ from $\ket \Xi$; alternatively, when they receive instructions to transmit their states to $\mathsf C$, they each send both of their qubits to $\mathsf C$ who performs four independent qubit measurements in the computational basis -- a \textit{quantum} measurement, but manifestly separable -- and from the outcome of these measurements, $\mathsf C$ is able to determine and report the resulting Bell state shared by $\mathsf A$ and $\mathsf B$ according to the table in Figure \ref{f:pironiostyle}. 

Once again, the entangled measurement has been displaced from $\mathsf C$ to other measurements: the measurements performed at $\mathsf {C_A}$ and $\mathsf {C_B}$, as $U^\dagger$ is entangling. One can further reinterpret the (local) entangling action of $U^\dagger$ at $\mathsf {C_A}$ and $\mathsf{C_B}$ as unentangled by encoding each of the two qubit-pair systems into four-level systems (ququarts) following the relabeling of \eqref{e:ququart}, which are sent to $\mathsf C$ in rounds where they are not measured separately at $\mathsf{C_A}$ and $\mathsf{C_B}$. In this case, no entangled measurement occurs anywhere in the model, though as previously noted this depends on a rather restrictive paradigm requiring four level systems that admit no interpretation as two subsystems, even while effectively encoding nontrivial dynamics of two separate qubits.

An interpretation of $\mathsf C$'s measurement according to the technical results of Ref.~\cite{bancal:2018} is given in supplementary material \cite{SM}. As a notable final point, the above model is compatible with a reversed layout in which $\mathsf{C_A}$ and $\mathsf{C_B}$ lie in the forward lightcone of $\mathsf C$.

\begin{figure}
\hspace{-1cm}\includegraphics[scale=1.5]{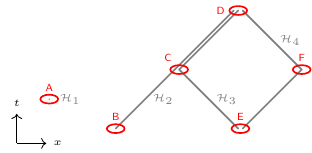}
\caption{Example of a general spacetime-ordered device-independent scenario. For an arbitrary collection of spacetime events at which measurements occur (vertices $\mathsf{A}-\mathsf{F}$), associate to each maximal-length timelike path a Hilbert space $\mathcal H_i$ indicating a possible route taken by quantum information. This includes zero-length paths at events spacelike separated from all others. Here the resultant state space is $\mathcal H_1\otimes \mathcal H_2 \otimes \mathcal H_3 \otimes \mathcal H_4$; a measurement at $\mathsf C$ for example acts on $\mathcal H_2 \otimes \mathcal H_3$, while $\mathsf B$ acts only on $\mathcal H_2$.}\label{f:spacetime}
\end{figure}

\medskip

\noindent \textit{General impossibility result.} I now generalize from scenario-specific schemes to an impossibility result for a broad class of measurement scenarios. First note Figure \ref{f:pironiostyle} can be reinterpreted as a \textit{localization protocol} (see, e.g., \cite{gisin:2024} Section IV.2): a method for performing a joint measurement in a distributed manner using ancillary entangled states and one-way classical transmission of local outcomes. Localization protocols, which have recently received renewed attention \cite{gisin:2024,pauwels:2025}, can be constructed to arbitrary accuracy for fully general joint measurements. One can construct
a procedure for removing an entangled measurement from any measurement scenario by modifying and adapting a localization protocol of Vaidman \cite{vaidman:2003}. This is formalized with the following result, which for simplicity is stated for a two-party/two-qubit projective measurement, but is generalizable: 

\medskip

\begin{figure}
\hspace{-2mm}\includegraphics[scale=0.65]{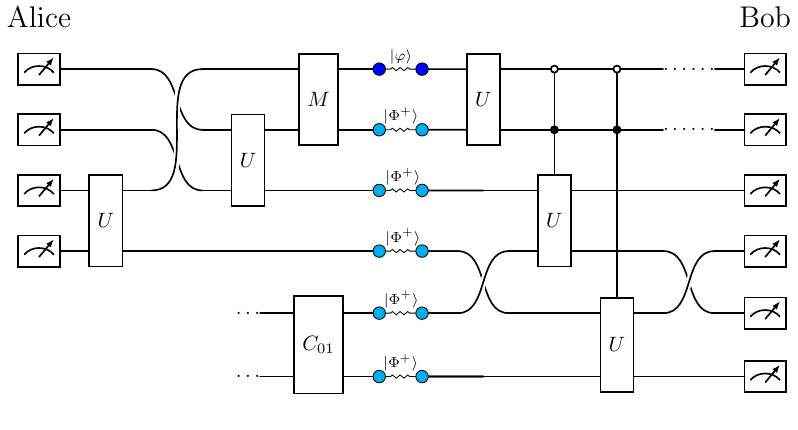}
\caption{Modification of the Vaidman protocol \cite{vaidman:2003} in which all BSMs are disambiguated as in Fig.~\ref{f:BSM}. The target state is $\ket \varphi$ in the top wire; the top four wires represent the first level of the protocol utilizing three ancillary copies of $\ket {\Phi^+}$. First, Bob uses the second wire to teleport his portion of $\ket \varphi$ to Alice (left side). Alice, without applying a teleportation correction -- which is unavailable in a factoring protocol -- applies a unitary $M$ whose construction depends on the specific choice of two-qubit measurement $\mathcal M$, then teleports the two-qubit state back to Bob (again without correction) in the 3rd and 4th wires. If Bob's top two measurements yield outcomes $(0,0)$, his computational basis measurements of the 3rd and 4th wires replicate $\mathcal M$ following the reasoning of the original protocol \cite{vaidman:2003}. If Bob observes instead $(0,1)$, the protocol proceeds to the second level where Bob teleports his two qubits back to Alice in a ``port" comprising wires 5 and 6, upon which Alice implements a port-dependent unitary correction $C_{01}$ followed by a process (not depicted) analogous to her level 1 manipulations. Importantly, one can employ the principle of deferred measurement to implement Bob's conditional-on-outcome-$(0,1)$ teleportation with a controlled-$U$ operation followed by the existing computational basis measurement of the top two wires. The full implementation of the protocol's second level requires two additional controlled teleportations corresponding to Bob's alternate outcomes $(1,0)$ and $(1,1)$ (not depicted; the controls would occur at the dots on the top two wires); this entails four additional wires for Bob to teleport the two-qubit state to Alice through two alternate ports, and six more wires for Alice to teleport all three ports back to Bob. An interesting feature of the Vaidman protocol is that Alice must perform all the teleportations in all ports since she is not privy to information about which outcomes were observed by Bob in the first level; Bob will ignore all but one port. The protocol is iterated to higher levels until the desired probability of success is achieved.}\label{f:vaidman} 
\end{figure}

\medskip

\noindent \textbf{Proposition.} Let $\mathcal M$ be an arbitrary projective quantum measurement on $\mathbb C^2\otimes \mathbb C^2$. For any $\epsilon>0$, there exists a positive $n$ and two unitaries $V$ and $W$ each acting on $\mathbb C^2 \otimes \mathbb C^{2n}$ such that for any state ${\ket \varphi \in\mathbb C^2 \otimes \mathbb C^2}$, computational basis measurements performed on ${V\otimes W\big[\ket \varphi_{VW}\otimes(\ket{\Phi^+}_{VW})^{\otimes n}\big]}$ replicate the effect of $\mathcal M$ on $\ket \varphi$ with probability at least $1-\epsilon$.

\medskip

\noindent A scheme that applies the (possibly entangled) measurement $\mathcal M$ to $\ket \varphi$ at some spacetime event can thus be replicated to any desired degree of accuracy with another scheme that appends $n$ copies of $\ket{\Phi^+}$ to $\ket \varphi$, noting $\ket \varphi$ may in general be entangled with other systems at other locations, and applies $V\otimes W$ as in the proposition to prepare the new underlying state. Then, $\mathcal M$ is replaced by computational basis measurements; at other spacetime events where $\ket \varphi$ was previously measured (in part or in whole), apply $V^\dagger$ and/or $W^\dagger$ to the new underlying state prior to measurement. This may cause these other measurements to newly become entangled measurements, though as previously discussed a local relabeling \`a la \eqref{e:ququart} allows interpretation of the state in terms of high-dimensional qudits on which $V^\dagger$ and $W^\dagger$ are not entangling. The procedure can be applied to any spacetime-ordered device-independent scenario of the general type illustrated in Figure \ref{f:spacetime}.

\medskip 

\noindent \textit{Proof}. The protocol of Vaidman -- see Ref.~\cite{vaidman:2003}; also \cite{pauwels:2025} Section III.A contains a good recent exposition -- involves multiple rounds of ``blind" teleportation of $\ket \varphi$ between the parties Alice and Bob seeking to implement $\mathcal M$. Depending on the outcomes of Bob's teleportation BSMs, Bob either achieves a ``success" condition indicating he can faithfully implement $\mathcal M$ locally, or he proceeds to a subsequent level of the protocol involving another round of teleporting the state back and forth. The probability of success can be made arbitrarily close to 1 with a sufficient number of rounds. For our purposes, two key tools -- the disambiguation of BSMs as in Fig.~\ref{f:BSM}, along with the principle of deferred measurement (\cite{NC} Section 4.4) -- allow construction of a variant of the Vaidman protocol in which all measurements are in the computational basis; the construction is demonstrated in Figure \ref{f:vaidman}. The unitaries $V$ and $W$ comprise all of the gates implemented by Alice and Bob, respectively, prior to measurement. \hfill$\Box$

\medskip

Many generalizations of the above result are possible. Higher-dimensional $\ket \varphi$ states can be encoded in multiple qubits for a parallelized version of the protocol in which each teleportation is replaced with multiple teleportations, and nonprojective measurements can be considered using Peres's routine for replacing POVMs with projective measurements \cite{PeresQT}, which is well suited for device-independent scenarios \cite{bierhorst:2023}. Multi-party entangled measurements can be tackled with Vaidman's generalization of the localization protocol to multiple parties \cite{vaidman:2003} and a corresponding Figure \ref{f:vaidman}-style implementation. And while the current formulation presupposes that a given portion of a state is only measured once in any given experimental trial (though the measurement can occur at different spacetime events in different trials), sequential measurements on postmeasurement states can be incorporated by leveraging the fact that the post-$\mathcal M$ state of $\ket \varphi$ is preserved on Bob's side of Figure \ref{f:vaidman}. Finally, if one allows for higher dimensional qudits and adopts a strict interpretation whereby they admit no entangled measurements when measured as single systems, the procedure can be iterated to remove entangled measurements from a scheme one-by-one. The impossibility result thereby extends to a very general class of scenarios, casting doubt on whether it is possible to design a scheme whose behavior requires a given one of its measurements to be entangled, at least according to current conceptions of device independence.

The models in this work all require a certain degree of complexity; this suggests an unavoidable tradeoff when displacing entanglement from measurements to states. Protocols involving entangled measurements can thus be re-conceptualized as device-independently certifying other more fundamental resources, such as (for example) minimal Hilbert space dimension, for which a connection to entangled measurements has been recognized previously \cite{cong:2017}. A particularly intriguing connection to explore is a tradeoff between entanglement of a measurement and the recently introduced quantity of localization cost \cite{pauwels:2025}. Nonetheless, the intractability of certifying an entangled measurement device-independently poses a challenge for how best to witness this important primitive in quantum networks. Even in an experiment flawlessly performing a basic entanglement swap and maximally violating CHSH inequalities on the post-swap states, one cannot certify in a fully device-independent manner that the swapping measurement is entangled, because appropriate measurements on a more complicated underlying state can replicate the complete behavior with the apparant swapping measurement comprising only independent qubit measurements. This indicates that some device-dependent assumptions, such as independent sources or restricted Hilbert space dimension \cite{renou:2018,bierhorst:2023,vertesi:2011}, are indispensable to certifying an entangled measurement.

\medskip

\begin{acknowledgments}
The author thanks Nicholas Fore for assistance in designing Figures 1, 3, and 4. This work was partially supported by NSF Award No.~2328800.
\end{acknowledgments}

\bibliographystyle{unsrt}
\bibliography{metabib}

%\end{document}
%What follows is the Supplementary Material section.

\clearpage

\pagestyle{empty}

\setcounter{equation}{0}

% Prepend an "S" to the numbering (e.g., Fig. S1, Eq. S1)
\renewcommand{\theequation}{S\arabic{equation}}
\renewcommand{\thefigure}{S\arabic{figure}}
\renewcommand{\thetable}{S\arabic{table}}

\onecolumngrid
\begin{center}
    \textbf{\large Supplementary Material: Intractability of Witnessing an Entangled Measurement Device Independently}
\end{center}

\PBc{This section reproduces three paragraphs that previously were located immediately preceding the subsection in the main text titled ``\textit{General Impossibility Result}''. All deviations from a verbatim reproduction are highlighted. The longer highlighted passage in the third paragraph is adapted from a different stricken portion of the main text on the same page.}
 
It is informative to apply the technical results of Ref.~\cite{bancal:2018} to the \PBs{above model}\PB{model that was described in conjunction with Figure \ref{f:pironiostyle} in the main text}, to understand the sense in which the \PB{model's }measurement at $\mathsf C$ is certified to be a Bell state measurement according to the framework of \cite{bancal:2018}. The notion is formulated mathematically as equation (6) of \cite{bancal:2018}, which with a relabelling to match \PBs{our model we}\PB{the present paper's notation one }can write as: 
\begin{equation}\label{e:bancal6}
\mathcal M \circ ( \Lambda^{(1)} \otimes \Lambda^{(2)} \otimes I_\mathsf{AB})[\rho_{\mathsf{C_AC_B}}\otimes \tau]=\bar{ \mathcal M} [\tau].
\end{equation}
On the left side of \eqref{e:bancal6}, $\mathcal M$ represents the mathematical description of the actual measurement in the experiment -- \PBs{here, }\PB{which for the model in the main text is }four computational-basis qubit measurements at $\mathsf C$.  Then $\rho_{\mathsf{C_AC_B}}$ is ``the state of the source used in practice,'' which is the reduced state of the four middle qubits of \PB{the state }$\ket \Xi$ \PB{defined in the main text}. The idea of \eqref{e:bancal6} is when these components are composed with hypothetical channels $\Lambda^{(1)}$ and $\Lambda^{(2)}$ and applied to two qubits of a hypothetical state $\tau$, the result is mathematically equivalent to a true BSM, written above as $\bar{ \mathcal M}$, applied to this same state $\tau$. More formally, $\Lambda^{(1)}:\mathcal H_{\mathsf{C_A}}\otimes \mathbb C^2_{\mathsf{T_A}} \to \mathcal H_{\mathsf{C_A}}$ and $\Lambda^{(2)}:\mathcal H_{\mathsf{C_B}}\otimes \mathbb C^2_{\mathsf{T_B}} \to \mathcal H_{\mathsf{C_A}}$ are completely postitive trace preserving maps, $\tau$ is a state in $\mathbb C^2_{\mathsf {T_A}}\otimes\mathbb C^2_{\mathsf {T_B}}\otimes \mathcal H_{\mathsf A}\otimes \mathcal H_{\mathsf B}$, and $\bar{ \mathcal M}$ is a true Bell state measurement on the first two qubits of $\tau$ while acting as identity on $\mathcal H_{\mathsf A}\otimes \mathcal H_{\mathsf B}$.

\PBs{For our model, }\PB{For the model in the main text, the $\mathsf{C_A}$ party measures two-qubit observables $U (S^{\vec r}_{\mathsf {C^1_A}} \otimes I_{\mathsf {C^2_A}}) U^\dagger$, where $S^{\vec r}$ is the settings-dependent single-qubit Bloch sphere observable in the honest/ideal implementation of two $\ket {\Phi^+}$ states (represented in Figure 3 of the main text), and $U=U_{\mathsf {C^1_AC^2_A}}$ is as defined in Eq.~\eqref{e:Udef}. The party at $\mathsf{C_B}$ acts similarly. In this scenario, }one can realize expression \eqref{e:bancal6} with the following maps: $\Lambda^{(1)}$ is $U^\dagger_{\mathsf{C^1_AC^2_A}}\otimes I_{\mathsf {T_A}}$, followed by tracing out over $\mathsf{C^1_A}$, followed by $U_{\mathsf{C^2_AT_A}}$; analogously, $\Lambda^{(2)}$ is $U^\dagger_{\mathsf{C^1_BC^2_B}}\otimes I_{\mathsf {T_A}}$, then $\text{tr}_{\mathsf {C^1_B}}[\cdot]$,  then $U_{\mathsf{C^2_BT_B}}$. The first step restores the state on $\mathcal H_{\mathsf{C^2_A}} \otimes \mathcal H_{\mathsf{C^2_B}}$ to $\ket {\Phi^+}$, while the last allows the computational basis measurements of $\mathcal M$ to serve as two BSMs on $\mathcal H_{\mathsf{C^2_A}} \otimes \mathcal H_{\mathsf{T_A}}$ and $\mathcal H_{\mathsf{C^2_B}} \otimes \mathcal H_{\mathsf{T_B}}$. Crucially though, the $\Lambda^{(1)}/\Lambda^{(2)}$ maps are only virtual elements not corresponding to anything in the experiment. The actual measurement $\mathcal M$ performed at $C$ would be able to effect an entangled measurment on the two qubits of a $\tau$-like state only after introducing an ancilary entangled state $\ket {\Phi^+}_{\mathsf{C^2_AC^2_B}}$, along with two machines to perform the $\Lambda^{(1)}/\Lambda^{(2)}$ tasks that apply the entangling operations $U_{\mathsf{C^2_AT_A}}$ and  $U_{\mathsf{C^2_BT_B}}$. Such elements could be pre-appended to any four-qubit unentangled measurement to similar effect; the essence of the entangling character of any resulting measurement lies in the pre-appended elements.

A key feature of the above constructions arises from the ``junk" states that can accompany self-testing isometries (see \cite{supic:2020} equations (10) and (11), for example) -- such as the isometries of Eq.~(14) in \cite{bancal:2018}, where tracing out over extraneous states prior to computing fidelity is not made explicit. A ``junk" state entangled over the $\mathsf{C_A}$/$\mathsf {C_B}$ partition (such as $\ket {\Phi^+}_{\mathsf {C^2_AC^2_B}}$) undermines the adaptation of these isometries to conditions indicating entangled measurements (equations (5)-(7) in \cite{bancal:2018}). In particular, it prevents the incorporation of $\rho_{\mathsf{C_AC_B}}$ into factoring $\Lambda^{(1)}/\Lambda^{(2)}$ maps as proposed in the caption of Figure 1 of \cite{bancal:2018}. Such a strengthening is however possible under the additional device-dependent assumption of independent sources discussed in \cite{bancal:2018}, which is the assumption made by the contemporaneous work Ref.~\cite{renou:2018}.

\vspace{5mm}

\begin{center}
\textbf{References}
\end{center}

\begin{enumerate}\setlength\itemsep{-1mm}
    \item[{[3]}] M.-O.~Renou, J.~Kaniewski, and N.~Brunner. Self-testing entangled measurements in quantum networks. \textit{Phys.~Rev.~Lett.}, 121:250507, Dec 2018. 
    \item[{[9]}] J.-D. Bancal, N.~Sangouard, and P.~Secatski. Noise-resistant device-independent certification of Bell state measurements. \textit{Phys. Rev. Lett.}, 121:250506, Dec 2018.
    \item[{[11]}] I.~\v{S}upi\'c and J.~Bowles. Self-testing of quantum systems: a review. \textit{Quantum}, 4:337, September 2020.
\end{enumerate}

\end{document}